\documentclass[aps,pra,reprint,superscriptaddress,showpacs,showkeys
]{revtex4-2}
\usepackage{txfonts}
\usepackage{color}

\usepackage{graphicx}
\usepackage{xcolor}
\usepackage{siunitx}

\begin{document}

\title{Single-shot Stern-Gerlach magnetic gradiometer with an expanding cloud of cold cesium atoms}

\author{Katja Gosar}
\affiliation{Jo\v{z}ef Stefan Institute, Jamova 39, SI-1000 Ljubljana, Slovenia}
\affiliation{Faculty of Mathematics and Physics, University of Ljubljana, Jadranska 19, SI-1000 Ljubljana, Slovenia}

\author{Tina Arh}
\affiliation{Jo\v{z}ef Stefan Institute, Jamova 39, SI-1000 Ljubljana, Slovenia}
\affiliation{Faculty of Mathematics and Physics, University of Ljubljana, Jadranska 19, SI-1000 Ljubljana, Slovenia}

\author{Tadej Me\v{z}nar\v{s}i\v{c}}
\affiliation{Jo\v{z}ef Stefan Institute, Jamova 39, SI-1000 Ljubljana, Slovenia}
\affiliation{Faculty of Mathematics and Physics, University of Ljubljana, Jadranska 19, SI-1000 Ljubljana, Slovenia}

\author{Ivan Kvasi\v{c}}
\affiliation{Jo\v{z}ef Stefan Institute, Jamova 39, SI-1000 Ljubljana, Slovenia}

\author{Du\v{s}an Ponikvar}
\affiliation{Jo\v{z}ef Stefan Institute, Jamova 39, SI-1000 Ljubljana, Slovenia}
\affiliation{Faculty of Mathematics and Physics, University of Ljubljana, Jadranska 19, SI-1000 Ljubljana, Slovenia}

\author{Toma\v{z} Apih}
\affiliation{Jo\v{z}ef Stefan Institute, Jamova 39, SI-1000 Ljubljana, Slovenia}

\author{Rainer Kaltenbaek}
\affiliation{Faculty of Mathematics and Physics, University of Ljubljana, Jadranska 19, SI-1000 Ljubljana, Slovenia}

\author{Rok \v{Z}itko}
\affiliation{Jo\v{z}ef Stefan Institute, Jamova 39, SI-1000 Ljubljana, Slovenia}
\affiliation{Faculty of Mathematics and Physics, University of Ljubljana, Jadranska 19, SI-1000 Ljubljana, Slovenia}

\author{Erik Zupani\v{c}}
\affiliation{Jo\v{z}ef Stefan Institute, Jamova 39, SI-1000 Ljubljana, Slovenia}

\author{Samo Begu\v{s}}
\affiliation{Faculty of Electrical Engineering, University of Ljubljana, Tr\v{z}a\v{s}ka cesta 25, SI-1000 Ljubljana, Slovenia}

\author{Peter Jegli\v{c}}
\email[]{peter.jeglic@ijs.si}
\affiliation{Jo\v{z}ef Stefan Institute, Jamova 39, SI-1000 Ljubljana, Slovenia}

\date{\today}

\begin{abstract}
We combine the Ramsey interferometry protocol, the Stern-Gerlach detection scheme, and the use of elongated geometry
of a cloud of fully
polarized cold cesium atoms to measure the selected component of the magnetic field gradient along the atomic cloud in a single shot.
In contrast to the standard method where the precession of two spatially separated atomic clouds is simultaneously measured to extract their phase difference, which is proportional to the magnetic field gradient, we here demonstrate a gradiometer using a 
single image of an expanding atomic cloud with the phase difference imprinted along the cloud.
Using resonant radio-frequency pulses and Stern-Gerlach imaging, we first demonstrate nutation and Larmor precession of atomic magnetization in an applied magnetic field. 
Next, we let the cold atom cloud expand in one dimension and apply the protocol for measuring the magnetic field gradient.
The resolution of our single-shot gradiometer is not limited by thermal motion of atoms and has an estimated absolute accuracy below $\pm0.2$~mG/cm ($\pm20$~nT/cm).

\end{abstract}


\maketitle

\section{Introduction}
Atomic magnetometers are among the most precise devices for measuring magnetic fields \cite{Grosz_2017,Mitchell_2020}.
The magnetic field magnitude is determined by the Larmor precession frequency of spins that is proportional to the field they are subjected to.
Centimeter-sized alkali-vapor magnetometers can be applied to measure the magnetic field either as a vector
quantity or as a scalar magnitude, depending on the method.
They can reach magnetic field sensitivity as high as $160$~aT/Hz$^{1/2}$ \cite{Dang_2010}.

\par

Ultracold atoms are very suitable for high-precision measurements due to their long lifetimes and small Doppler broadening \cite{Sycz_2018}.
The sensitivity of cold-atom magnetometers does not reach that of the best alkali-vapor devices because of the small size of atom clouds at comparable densities, but they are suitable for measurements with a high spatial resolution.
They can reach $8.3$~pT/Hz$^{1/2}$ magnetic field sensitivity on a $\sim10$~$\mu$m scale \cite{Vengalattore_2007}, and they can detect magnetic-field inhomogeneities down to $200$~nT/cm \cite{Koschorreck_2011}.
Typically, cold-atom magnetometers are based on the same signal detection technique as room-temperature atomic-vapor
magnetometers, i.e., the Faraday rotation \cite{Franke-Arnold_2001,Labeyrie_2001,Takeuchi_2006,Koschorreck_2010, Wojciechowski_2010, Behbood_2013, Eliasson_2019,Cohen_2019}.
This is, however, not the only possibility for detecting the spin precession in cold-atom clouds.
Various techniques have been developed, including state-selective phase-contrast imaging \cite{Vengalattore_2007,Higbie_2005} and state-selective absorption imaging \cite{Koschorreck_2011}. 
Finally, the projection of the magnetization can also be measured through the populations of Zeeman sublevels using
the Stern-Gerlach method, in combination with using the Ramsey sequence to control the precession time
\cite{Eto_2013_APE,Eto_2013_PRA,Sadgrove_2013,Wood_2015}. This is the approach adopted in this work.

\par

A basic gradiometer consists of two magnetometer probes separated in space,
and the magnetic gradient is obtained by differentiating their outputs.
Magnetic gradiometry using two clouds of cold atoms was demonstrated in Ref. \onlinecite{Wood_2015}.
The experimental setup allowed the control of cloud positions, thereby enabling the measurement of the complete magnetic field gradient tensor.
In inhomogeneous magnetic field the precession frequency is position dependent.
For uniform initial phase, the phase difference between the clouds at positions $\vec{r_1}$ and $\vec{r_2}$
accumulates with time as:
\begin{equation}
	\Delta\phi(t) = \gamma \left(B(\vec{r_2})  - B(\vec{r_1})  \right) t \approx  \gamma t  \, (\vec{r_2}- \vec{r_1}) \cdot \nabla B.
\label{Deltaphi}	
\end{equation}
Here, $\gamma$ denotes the gyromagnetic ratio.
The phase difference is therefore proportional to the spatial derivative of the magnetic field strength $B$ along the direction connecting both probes.
Traditionally, a series of measurements with incrementing interrogation (precession) times is required to determine the time evolution of the phase difference.
Special care has to be taken to properly count the integer multiples of $2\pi$ \cite{Wood_2015}.

\par
Here we demonstrate a method for measuring the selected component of the magnetic field gradient with a single shot,
using only one elongated atom cloud instead of two spatially separated clouds.
Specifically, we measure the spatial profile of the magnetic field through the spatial dependence of the phase
difference $\Delta\phi(x,t)$ along the cloud elongated in the $x$-direction, see Fig.~\ref{fig1}(a).
For a given precession time $t$, the magnetic field gradient causes the accumulated precession phase to have a
continuous variation along the atom cloud.
We choose the magnetic field $B_0$ to be oriented in the $x$-direction.
In this case, we can approximate the position dependence of the magnetic field magnitude with $B(x) \approx B_0 + ({ \partial B_x}/{ \partial x}) x$.
Here, we neglect contributions from magnetic field gradients ${\partial B_y}/{ \partial x}$ and ${\partial B_z}/{ \partial x}$, which may result in small magnetic-field components perpendicular to the dominant $B_0$.
The magnetization of the atoms in the cloud undergoes a Larmor precession in the magnetic field.
Therefore, its $y$-projection can be written as
\begin{equation}
M_y(x, t) = M_0 \cos \left( \gamma B_0 t + \gamma \frac{ \partial B_x}{ \partial x} x  t  \right).
\label{My}
\end{equation}
Here we assume that, at $t=0$, the whole cloud is fully polarized along the $y$-direction with initial magnetization $M_0$.
As we show below, a single experimental run with only one Stern-Gerlach image of the atom cloud is sufficient to obtain $M_y(x)$ at the selected precession time $t$, 
allowing an unambiguous extraction of the magnetic field gradient ${\partial B_x}/{ \partial x}$.
If the magnetic field $B_0$ is oriented in other directions, additional components of magnetic-field-gradient tensor can be determined.
For example, to determine  ${\partial B_y}/{ \partial x}$, the magnetic field $B_0$ has to be in the $y$-direction.

\par

According to Eq.~(\ref{My}), nonzero components of the magnetic field gradient along the cloud cause a helical or "corkscrew" spatial dependence of the magnetization direction that can also be observed in Bose-Einstein condensates \cite{Higbie_2005,Eto_2013_APE}.
It is important to note that the time-fluctuations of the magnetic field $B_0$ and any non-compensated homogeneous
external fields contribute only to the overall phase in $M_y(x,t)$.
In elongated condensates, the phase difference can also become spatially dependent in the presence of inhomogeneous internal magnetic fields caused by the spatially modulated structure of spin domains \cite{Sadler_2006,Vengalattore_2008, Kronjaeger_2010, Eto_2014}, presumably induced by long-range dipole interaction.
In contrast, recent non-destructive Faraday-rotation experiments showed no spontaneous domain formation in a tightly confined low-density $^{87}$Rb condensate \cite{Palacios_2018}.
In cold atom clouds, which have even lower densities, these effects can be neglected.

\par

In this work, we focus on the detection of magnetic field gradients originating from external sources.
A related technique is described in Ref. \onlinecite{Koschorreck_2011}, where an elongated but non-expanding cloud of $^{87}$Rb cold atoms is polarized with a pump beam pulse and the projection of the magnetization is detected with state-selective imaging.
In our $^{133}$Cs experiment, the magnetization of the expanding cloud is already fully polarized along the applied magnetic field, and we start the precession with a pulse of a radio-frequency (RF) magnetic field.
To measure $M_y(x,t)$ we perform the Stern-Gerlach imaging (Fig.~\ref{fig1}), where we apply a magnetic field gradient to separate the atoms in different Zeeman sublevels and calculate the magnetization from the atom populations in each sublevel.
After taking into account the effect of the cloud expansion during the protocol for measuring the magnetic field gradient, we reach an absolute accuracy below $\pm0.2$~mG/cm ($\pm20$~nT/cm) in a single shot.
The sensitivity of magnetic field gradient is enhanced by an order of magnitude  compared to Ref.~\onlinecite{Koschorreck_2011}, where they had $\sim5$ times more atoms in $\sim10$ times more elongated cloud, albeit at $\sim20$ times higher temperatures, which only allowed much shorter interrogation times (below 1~ms).
As discussed below, the resolution of our gradiometer is not limited by thermal motion (diffusion) of atoms since their in-trap velocity distribution is mapped into well-defined atom trajectories during the cloud expansion.

\section{Experiment}

\begin{figure} [t!]
\includegraphics[width=1.0\linewidth]{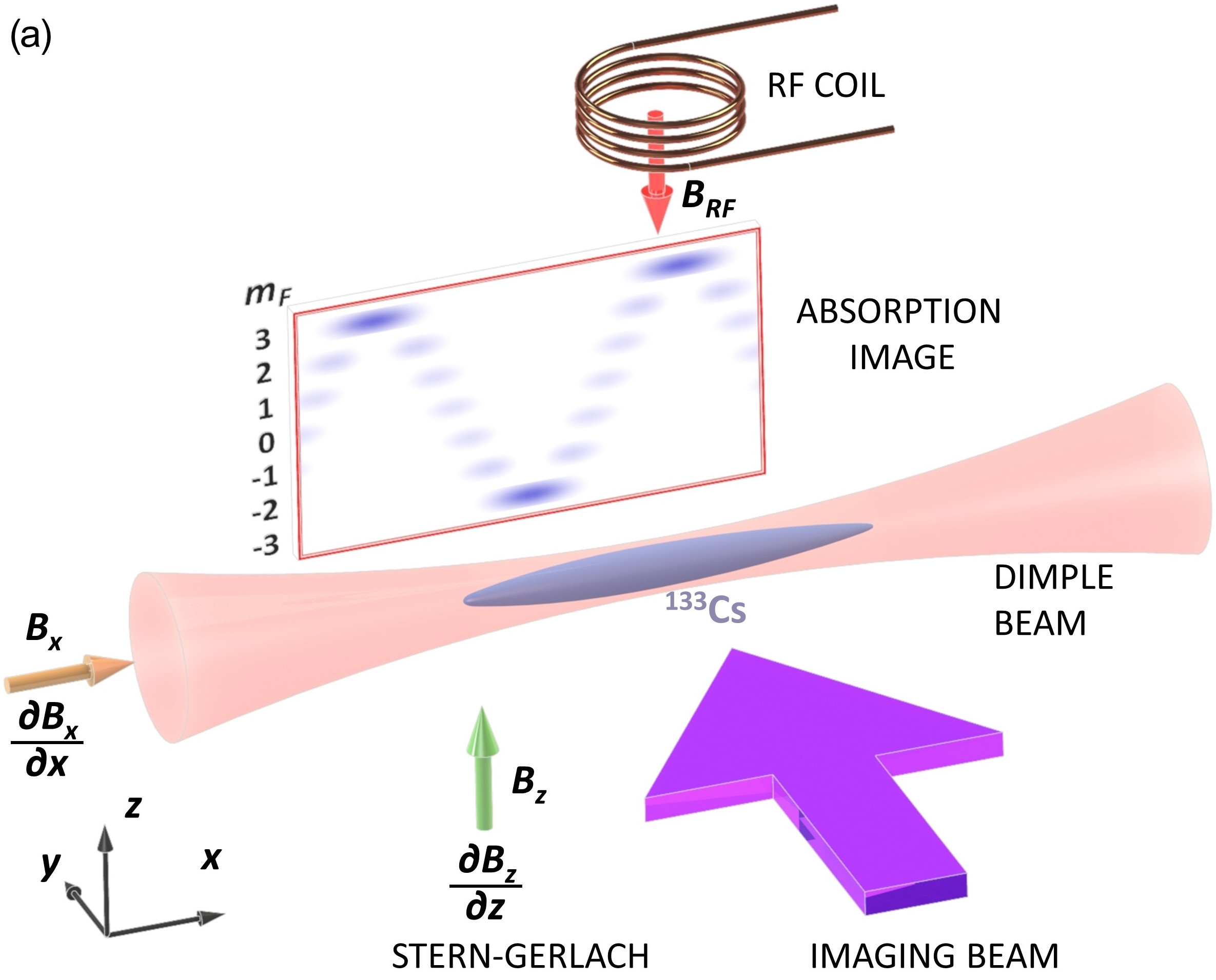}
\includegraphics[width=1.0\linewidth]{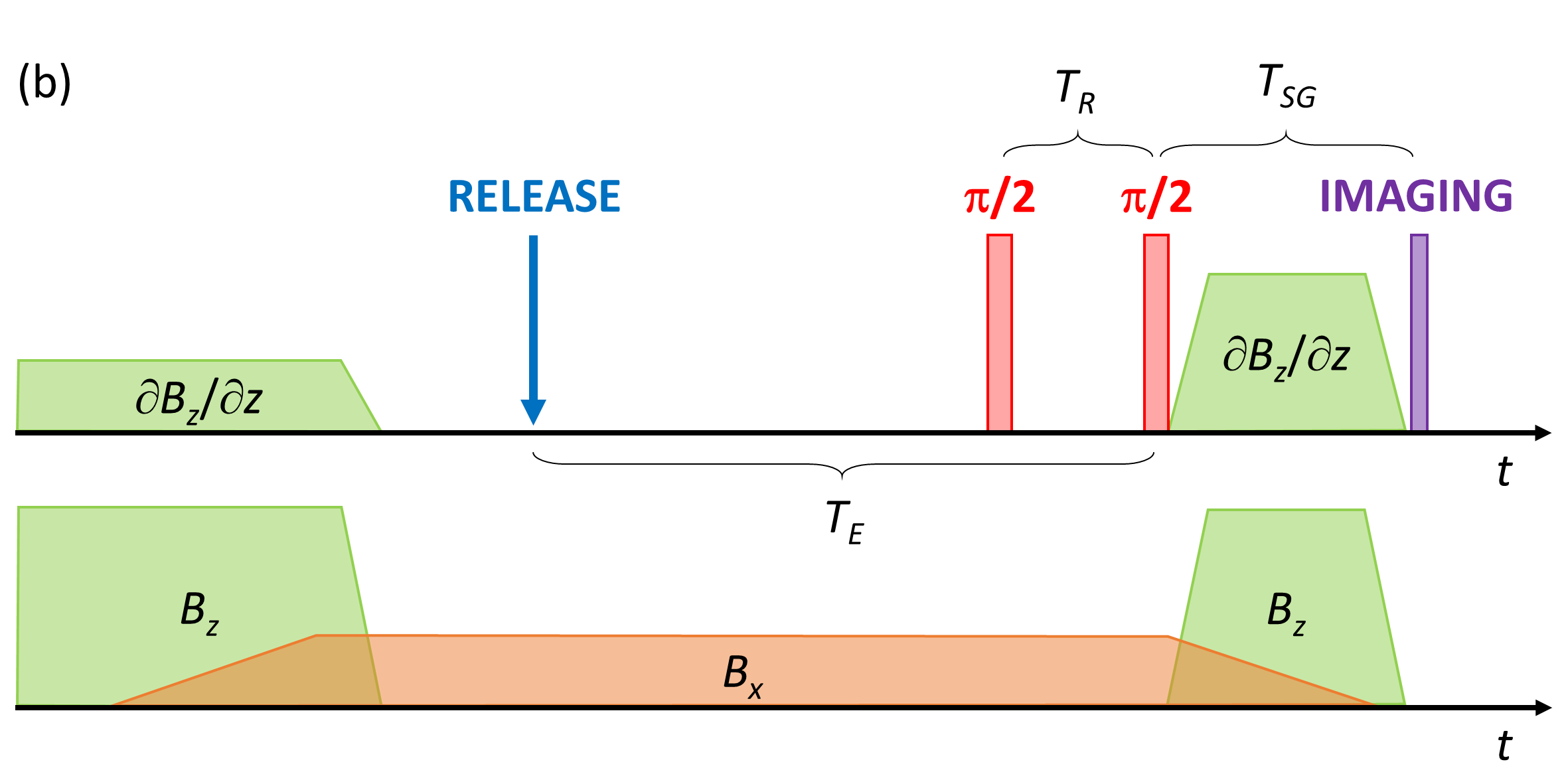}
\caption{(a) Schematic illustration of the magnetic gradiometer showing an elongated $^{133}$Cs cold atom cloud expanded along the dimple beam.
(b) The experimental sequence for observing position-dependent Larmor precession of magnetization, caused by the ${\partial B_x}/{ \partial x}$ component of the magnetic field gradient.
First, the magnetic field is switched from the $z$- to the $x$-direction, then the atom cloud is released to expand along the beam.
Next, we use the Ramsey sequence composed of two $\pi/2$ RF pulses that are separated by the interrogation time $T_R$. 
Finally, the absorption image is taken after the Stern-Gerlach separation of $m_F$-state populations in the applied magnetic field gradient ${\partial B_z}/{ \partial z}$.
}
\label{fig1}
\end{figure}

Most cold atom magnetometers and gradiometers are based on $^{87}$Rb atoms in $F = 1$ or $F = 2$ hyperfine states.
Here, we use  $^{133}$Cs atoms in their $F=3$ ground state; initially, they are fully polarized along the $x$-direction.
We prepare cold $^{133}$Cs atoms by laser cooling with a standard procedure described in detail in Ref.
\onlinecite{Meznarsic_2019}, including the transfer of fully polarized atoms in the $(F=3,m_F=3)$ state from a large
dipole trap to a small dimple trap with trap frequencies $2\pi \times(20,120,120)$~Hz, followed by
further evaporative cooling for $100$~ms.
The resulting cloud in the dimple trap typically consists of $2\times10^5$ cesium atoms at $T=1.29\pm0.02$~$\mu$K, 
with the $1/e$ radii of $\sigma_{x0}=69\pm2$~$\mu$m and $\sigma_{y0} \sim \sigma_{z0}=12\pm2$~$\mu$m.
To create an elongated atom cloud one of the dimple beams is turned off and the cloud starts expanding in the
$x$-direction, along the remaining beam  (Fig.~\ref{fig1}).
As shown below, a regime of linear-in-time expansion is reached after $\sim20$~ms of time-of-flight (TOF).
At $40$~ms of total expansion time, the cloud extends over an $1/e$ length of $\sigma_x=366\pm3$~$\mu$m.

\par

Immediately after the evaporation, we turn off the quadrupole coil producing a strong magnetic field gradient of $\partial B_z/\partial z=31.3$~G/cm used to levitate the cesium atoms (Fig.~\ref{fig1}(b)).
We also turn off the Helmholtz coil producing a homogeneous magnetic field $B_z=22$~G, which optimises the cesium scattering length during the evaporation.
At the same time we turn on the compensation coils that cancel out the components of magnetic field in the $y$- and $z$-directions and set the magnetic field in $x$-direction to a value of  $B_0=143$~mG.
That corresponds to a Larmor frequency of $\omega_0=\gamma B_0=2\pi\times 50$~kHz ($\gamma=350$~kHz/G).
Since the inductances of the quadrupole, Helmholtz and compensation coils are large (the switching times are in the order of several ms), we wait for $40$~ms for the magnetic fields to reach their final values.
During this period, the magnetization of the atoms, which are fully polarized in the $z$-direction, is adiabatically rotated to the $x$-direction since this process is slow compared to the Larmor precession \cite{Slichter}.
This approach is used to obtain the initial magnetization of the cold cesium atoms in all of the experiments presented below.

\par

To measure the Larmor precession of the magnetization, we apply the Ramsey sequence of RF pulses followed by the Stern-Gerlach measurement (Fig.~\ref{fig1}(b)).
The Ramsey sequence consists of two $\pi/2$ pulses, rotating the magnetization around the z-axis, separated by the interrogation time $T_R$.
With the first pulse, the magnetization is rotated from the $x$- to the $y$-direction, where it is then left to precess in the yz-plane until the second $\pi/2$ pulse is applied.
After the second pulse, the $y$-component of the instantaneous magnetization becomes the $x$-component, and the $z$-component stays unaffected.
Then we turn on a magnetic field $B_z=22$~G and a magnetic field gradient $\partial B_z/\partial z=100$~G/cm for Stern-Gerlach imaging.
Again, the magnetic field cannot reach its final value instantly.
Therefore, the magnetization of the atoms adiabatically follows the slow change of the orientation of the quantization axis.
This means that the $M_z$ component measured via Stern-Gerlach imaging is equal to the $M_y$ component at the moment right before the second $\pi/2$ pulse is applied.

\par

In the presence of a strong magnetic field gradient of $100$~G/cm, the atom cloud splits into separated clouds according to their $m_F$-states \cite{Eto_2013_APE,Eto_2013_PRA,Sadgrove_2013,Wood_2015}.
After $T_{SG}=10$~ms of Stern-Gerlach splitting, we take a standard absorption image of the separated clouds and calculate the expectation value of the spin $z$-component
\begin{equation}
\langle S_z \rangle = \frac{\sum_{m_F=-3}^{+3} m_F N_{m_F}} {\sum_{m_F=-3}^{+3} N_{m_F}},
\label{Sz}
\end{equation}
where $N_{m_F}$ is the atom number population in state $m_F$.
Similarly, we can calculate the variance of $S_z$ defined as 
\begin{equation}
\langle \Delta^2S_z \rangle = \frac{\sum_{m_F=-3}^{+3} m^2_F N_{m_F}} {\sum_{m_F=-3}^{+3} N_{m_F}}-\langle S_z \rangle^2.
\label{deltaSz}
\end{equation}
If the atoms are fully polarized and precess around the magnetic field perpendicular to their magnetization, the variance is equal to 0.5 when time averaged over one precession period \cite{Eto_2013_PRA}.
However, in the presence of decoherence, the time average of  $\langle \Delta^2S_z \rangle$ increases with the interrogation time $T_R$.
If the elongated cloud is oriented along the $x$-direction, the $z$-component of the spin and its variance will be functions of $x$: $\langle S_z(x) \rangle$ and $\langle \Delta^2S_z(x) \rangle$.

\section{Results}
\begin{figure} [t!]
\includegraphics[width=1.0\linewidth]{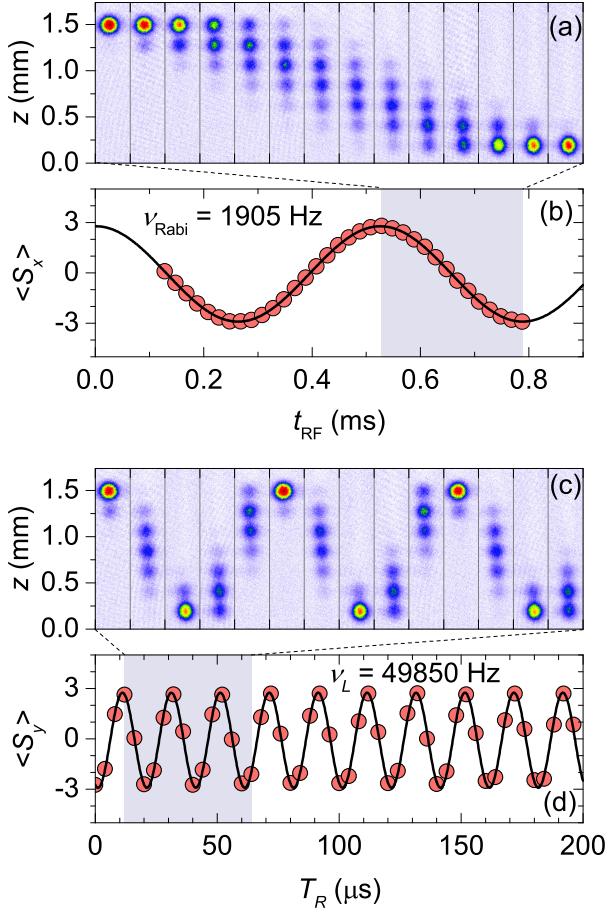}
\caption{(a) The Stern-Gerlach measurements showing separated atom clouds corresponding to the different $m_F$-state populations ranging from $m_F=-3$ (at the bottom) to $m_F=+3$ (at the top).
In the displayed absorption images the RF pulse length, $t_{RF}$, increases from 528~$\mu$s to 788~$\mu$s in steps of 20~$\mu$s.
(b) Oscillations of $\langle S_x \rangle$ as a function of $t_{RF}$ with $\nu_{\rm Rabi}=1905$~Hz (solid line).
(c) Absorption images showing the Larmor precession of magnetization, where the separation between two $\pi/2$ pulses, $T_R$, increases from 12~$\mu$s to 64~$\mu$s in steps of 4~$\mu$s.
(d) Oscillations of $\langle S_y \rangle$ as a function of $T_R$ with $\nu_L=49850$~Hz (solid line).
The red regions in (b) and (d) mark the experimental points shown in (a) and (c), respectively.
Each absorption image in (a) and (c) shows the area of $376$~$\mu$m~$\times$~$1738$~$\mu$m. 
In all these experiments $T_{SG}=10$~ms.
}
\label{fig2}
\end{figure}

Fig.~\ref{fig2}(a) shows the oscillations of the $m_F$-state populations of $^{133}$Cs atoms in the $F=3$ hyperfine state measured by the Stern-Gerlach method, applied immediately after one RF pulse of length $t_{\rm RF}$.
Using Eq.~(\ref{Sz}) we can calculate $\langle S_z\rangle$ as a function of $t_{\rm RF}$, which is equal to $\langle S_x \rangle$ immediately after the end of the RF pulse.
The nutation of magnetization observed as oscillations in $\langle S_x \rangle$ is displayed in Fig.~\ref{fig2}(b).
We obtain the Rabi frequency of $\nu_{\rm Rabi}=1905$~Hz and determine the $\pi/2$-pulse length to be $\sim130$~$\mu$s.

To observe the Larmor precession of magnetization, we apply the Ramsey sequence composed of two $\pi/2$-pulses followed by the Stern-Gerlach measurement.
By varying the time between the two pulses, the interrogation time $T_R$, we can observe fast oscillations of $m_F$-state populations as shown in  Fig.~\ref{fig2}(c).
Again, using Eq.~(\ref{Sz}) we can calculate $\langle S_z \rangle$, which is equal to $\langle S_y\rangle$ at the moment before the second $\pi/2$ pulse is applied.
The fast oscillations of $\langle S_y \rangle$ are displayed in Fig.~\ref{fig2}(d) from which we obtain the Larmor precession frequency of $\nu_L=49850$~Hz.
This allows for a more precise determination of the magnetic field magnitude, which is equal to $B_0=142.43$~mG.
A small mismatch between the values of $\nu_L$ and $\nu_{RF}$, and consequently imperfect $\pi/2$ pulse, 
results in small reduction in $\langle S_y\rangle$ oscillation amplitude, $S_0$, from the ideal value of $3$.

\begin{figure} [t!]
\includegraphics[width=1.0\linewidth]{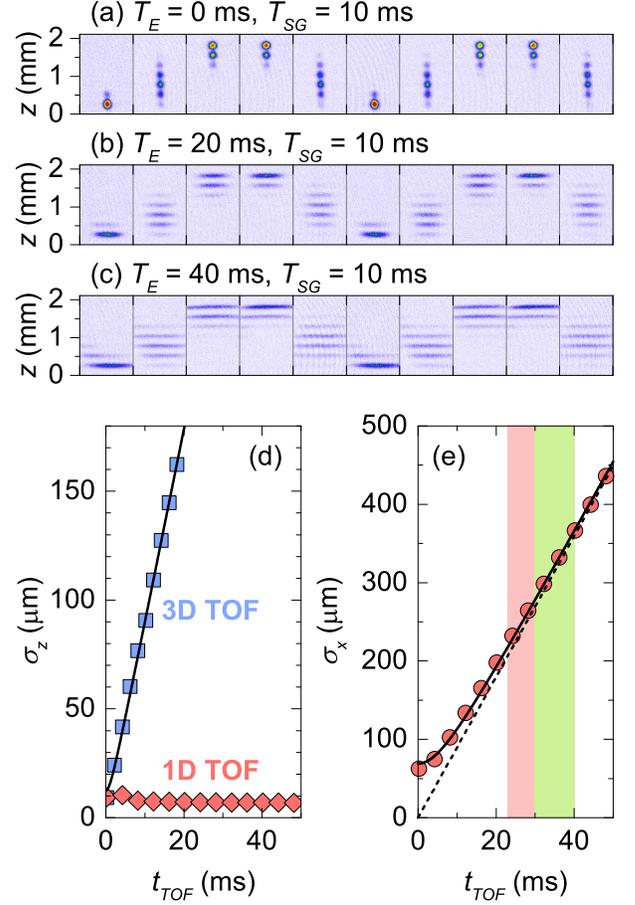}
\caption{Larmor precession of $m_F$-state populations for expansion times (a) $T_E=0$~ms, (b) $T_E=20$~ms and (c) $T_E=40$~ms.
Here the interrogation time $T_R$ runs from $0$~$\mu$s to $36$~$\mu$s in steps of $4$~$\mu$s ($T_{SG}=10$~ms).
The absorption images show the area of $1407$~$\mu$m~$\times$~$2110$~$\mu$m.
(d) The extracted $1/e$ widths $\sigma_z$ as a function of time-of-flight (TOF) during free-space expansion (blue squares, 3D TOF) and expansion along a single dimple beam (red diamonds, 1D TOF).
From the fit with the model $\sigma_z^2=\sigma_{z0}^2+k_BT_z/m\cdot t^2_{TOF}$, $\sigma_{z0}=12\pm2$~$\mu$m and $T_z=1.27\pm0.02$~$\mu$K are obtained.
(e) $\sigma_x$ for expansion along a dimple beam (red circles) together with the corresponding fit (solid line) using $\sigma_x^2=\sigma_{x0}^2+k_BT_x/m\cdot t^2_{TOF}$.
We obtain $\sigma_{x0}=69\pm2$~$\mu$m and $T_x=1.29\pm0.02$~$\mu$K.
The dashed line follows the linear-in-time expansion with $v_x=\sqrt{k_BT_x/m}=9.0\pm0.1$~mm/s. 
The red and green areas mark, respectively, the typical interrogation and Stern-Gerlach detection time intervals in our measurement protocol.
}
\label{fig3}
\end{figure}

\begin{figure} [t!]
\includegraphics[width=1.0\linewidth]{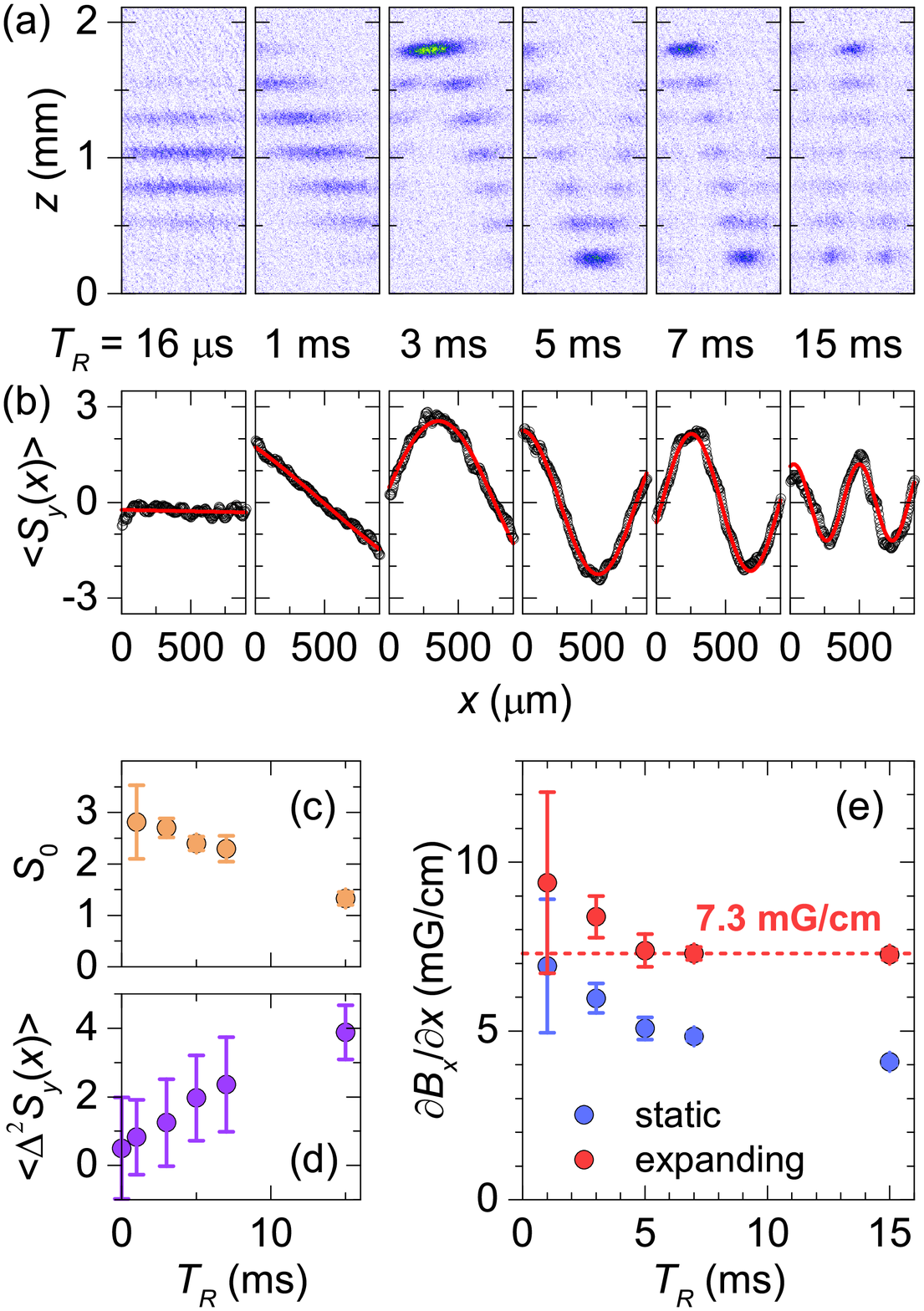}
\caption{(a) Absorption images of position dependent $m_F$-state populations for a range of interrogation times $T_R$.
The first image serves as a reference; it is taken with a short $T_R=16$~$\mu$s and shows position-independent populations of $m_F$-states.
The absorption images show the area of $914$~$\mu$m~$\times$~$2110$~$\mu$m ($T_E=30$~ms and $T_{SG}
=10$~ms).
(b) Extracted $\langle S_y(x)\rangle$ as a function of position $x$ for different $T_R$.
(c) The amplitude $S_0$ of the position dependent $\langle S_y(x)\rangle$ obtained from sinusoidal function fits (red lines in (b)).
(d) The variance $\langle \Delta^2S_y \rangle$, time averaged over one precession period.
(e) The extracted component ${\partial B_x}/{ \partial x}$ of the magnetic field gradient for two scenarios: (i) the static model (Eq.~(\ref{grad1}), blue circles) and (ii) the model taking into account the expansion of cold atom cloud (Eq.~(\ref{grad2}), red circles).
The error bars denote a standard deviation of magnetic field gradient obtained from 10 repetitions.
}
\label{fig4}
\end{figure}

In Figs.~\ref{fig3}(a,b,c) we show the Larmor precession of magnetization along the elongated cold atom clouds for different expansion times $T_E$.
For short interrogation times $T_R$, the $m_F$-state populations collectively oscillate in time for all $T_E$, meaning $\langle S_y(x)\rangle$ is almost independent of position $x$ along the cloud. 
However, for longer $T_R$, the $m_F$-state populations become space-dependent (Fig.~\ref{fig4}).
This is caused by the presence of the component ${\partial B_x}/{ \partial x}$ of the magnetic field gradient.
Fig.~\ref{fig4}(b) shows the position dependent $\langle S_y(x)\rangle$ for different values of $T_R$ and is a signature measurement of the presented magnetic gradiometer detection principle.

\begin{figure} [t!]
\includegraphics[width=0.6\linewidth]{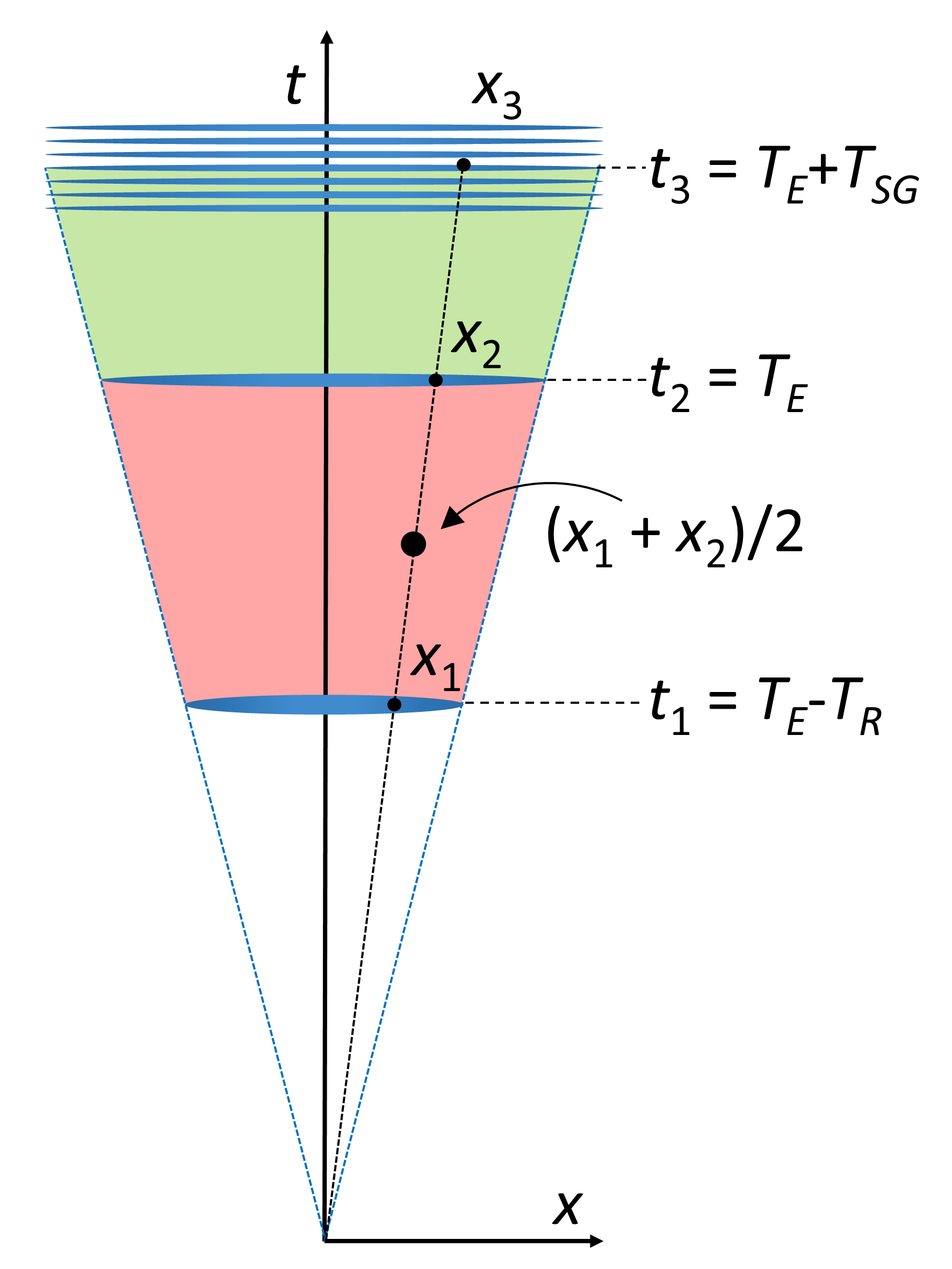}
\caption{Schematic illustration of cold atom cloud expansion. At time $t=0$, when the perpendicular dimple beam is turned off, the cloud starts expanding along the remaining dimple beam.
At $t_1$, the first $\pi/2$-pulse is applied and the magnetization starts precessing.
At $t_2$, the second $\pi/2$-pulse is applied and the Stern-Gerlach protocol starts (green area), ending at $t_3$, when the absorption image is taken.
The red area marks the time interval in which the position-dependent phase difference $\Delta\phi(x)$ is accumulated.
The atoms at the final position $x_3$ reflect the average magnetic field between positions $x_1$ and $x_2$. Because the magnetic field is linearly dependent on the position this is equal to the magnetic field at $(x_1 + x_2)/2$.
The corresponding time is exactly in the middle of the interrogation time interval at $(t_1+t_2)/2=T_E-T_R/2$. 
The renormalization factor in Eq.~(\ref{phi2}) follows directly from similar triangles: $(x_1+x_2)/2:x_3=(T_E-T_R/2):(T_E+T_{SG})$.
}
\label{fig5}
\end{figure}

Before we proceed with the evaluation of the component ${\partial B_x}/{ \partial x}$ of the magnetic field gradient, we first analyse the expansion of cold atom cloud in a dimple beam.
The initial cloud widths $\sigma_{z0}$ and $\sigma_{x0}$ are obtained from the analysis presented in Figs.~\ref{fig3}(d,e) by fitting the atom density profiles during both free-space expansion and expansion along the dimple beam. 
Notably, when $\sigma_x \gg \sigma_{x0}$, which is after $\sim20$~ms of expansion in a dimple beam, $\sigma_x$ starts increasing linearly in time with a velocity $v_x=\sqrt{k_BT_x/m}=9.0\pm0.1$~mm/s, where $k_B$ and $m$ are the Boltzmann constant and atomic mass of cesium, respectively.
This justifies a simple model of phase difference accumulation in expanding atom cloud schematically presented in Fig.~\ref{fig5}.

\par

For a nonexpanding cloud the phase difference $\Delta \phi_{\rm static}$ accumulated during the interrogation time $T_R$ can be derived directly from Eqs.~(\ref{Deltaphi}) and (\ref{My})
\begin{equation}
\Delta\phi_{\rm static}(x)=\gamma \frac{\partial B_x}{ \partial x} x T_R,
\label{phi1}
\end{equation}
which is proportional to ${\partial B_x}/{ \partial x}$ and $T_R$.
However, due to the effect of cloud expansion it can be shown using a simple geometric consideration that the phase difference $\Delta \phi_{\rm expanding}$ must be renormalized according to
\begin{equation}
\Delta\phi_{\rm expanding}(x)=\Delta\phi_{\rm static}(x)\cdot \frac{T_E-T_R/2}{T_E+T_{SG}}.
\label{phi2}
\end{equation}
The renormalization factor takes into account that during the interrogation time $T_R$ the atoms on average feel the magnetic field at the position $(x_1+x_2)/2$ (for details please refer to Fig.~\ref{fig5}).

\par

In Fig.~\ref{fig4}(b) we show the fits of experimental $\langle S_y(x)\rangle$ using Eq.~(\ref{My}) for a range of interrogation times $T_R$.
From each fit we obtain the wavelength $\lambda$ of helical spatial dependence of magnetization. From $\lambda$ we then calculate the component of the magnetic field gradient.
For nonexpanding case,
\begin{equation}
\left( \frac{\partial B_x}{ \partial x} \right)_{\rm static} =\frac{2\pi}{\gamma T_R \lambda}
\label{grad1},
\end{equation}
whereas for expanding cloud,
\begin{equation}
\left( \frac{\partial B_x}{ \partial x} \right)_{\rm expanding} =\frac{2\pi}{\gamma T_R \lambda}\cdot \frac{T_E+T_{SG}}{T_E-T_R/2}
\label{grad2}.
\end{equation}
The expression for the magnetic field gradient is independent of the atom-cloud temperature and that is one of the key results of our work.

The importance of taking into account the effect of expansion is best seen in Fig.~\ref{fig4}(e), where we plot and compare ${\partial B_x}/{ \partial x}$ obtained from Eqs.~(\ref{grad1}) and (\ref{grad2}).
Whereas for the static model (no expansion) the values of ${\partial B_x}/{ \partial x}$ decrease with increasing
interrogation time $T_R$, for the expanding-cloud model the extracted values become constant and their error decreases with increasing $T_R$ for $T_R\leq 15$~ms. 
For even longer $T_R$ the amplitude $S_0$ of modulated $\langle S_y(x)\rangle$ becomes substantially reduced (Fig.~\ref{fig4}(c)).
Additionally, its time averaged variance $\langle \Delta^2S_y \rangle$ increases (Fig.~\ref{fig4}(d)), meaning that
the decoherence of magnetization becomes important and reduces the sensor accuracy.
The reason behind this observation is mainly in our measurement protocol (Fig.~\ref{fig5}), where longer $T_R$ brings the interrogation interval closer to the point where the effects of finite initial size of atom cloud become relevant.
However, these mainly affect the decoherence of magnetization, but have only a very small impact on the evaluation of magnetic field gradient from Eq.~(\ref{grad2}). 
Even for $T_R=15$~ms the estimated systematic correction is below $3\%$.
Finally, for interrogation time $T_R=7$~ms we obtain ${\partial B_x}/{ \partial x}=7.3$~mG/cm with an estimated error
below $\pm0.2$~mG/cm in a single shot.
The measured magnetic field gradient is of external origin, most probably arising from ionic pumps surrounding our experimental chamber. 

\section{Discussion and Conclusions}

Using the described single-shot Stern-Gerlach magnetic gradiometer it is in principle possible to determine any
component of the complete magnetic-field-gradient tensor $\partial B_i/\partial j$, with $i,j=x,y,z$.
$\partial B_i$ can be selected by the direction of $B_0$ (in Fig.~\ref{fig1}, this is $B_x$), whereas $\partial j$ can be chosen by the orientation of the elongated cold atom cloud.
However, in order for the sensor accuracy to be similar for all measured $\partial B_i/\partial j$ components, the imaging should preferably be perpendicular to the elongated cloud and $B_{RF}$ perpendicular to $B_0$ (Fig.~\ref{fig1}).
The presented method has the potential for miniaturization using atom chips \cite{Keil_2016,Becker_2018}.
Such technology shortens the time needed to prepare cold atoms and facilitates the integration of laser beams and magnetic coils for producing the necessary magnetic fields.
For our method it is important that $B_{RF}$ and  $B_0$ are homogeneous.
If sufficient homogeneity cannot be achieved with integrated coils, external coils could be used with an atom chip.
This type of device could also easily be rotated in space to measure the magnetic-field gradients in an arbitrary direction.

\par

While the sensitivity and the accuracy of our device are comparable to or even surpass current proposals \cite{Vengalattore_2007,Koschorreck_2011}, it lacks temporal resolution.
The Stern-Gerlach detection of magnetization is destructive and $\sim13$~s are required to prepare a new elongated cloud of cold cesium atoms for each measurement.
Using $^{87}$Rb, a much faster production of cold atoms is possible and could reduce the temporal resolution well below $1$~s.
For example, in an all-optical $^{87}$Rb setup, rates on the order of $10^7$ cold atoms per second were reported \cite{Kinoshita_2005}.
In Ref.~\onlinecite{Rudolph_2015} atom-chip technology was used to achieve an even higher rate of about $10^8$ atoms per second.

\par

There are multiple ways to improve the sensitivity of the gradiometer presented here while retaining the same benefits and working principles.
The minimal measurable gradient is limited by the length of the cloud since the precision of our method decreases for longer wavelengths $\lambda$.
Therefore, a longer cloud would allow measurement of smaller gradients.
The other option is to use longer interrogation times $T_R$.
This improves the precision because $\lambda$ decreases for the same gradient.
However, $T_R$ cannot be increased indefinitely, because the signal amplitude $S_0$ decreases due to decoherence.
It would immediately be possible to allow longer $T_R$, if one extends the magnetization coherence time. 
This depends on satisfying the condition $\sigma_x \gg \sigma_{x0}$.
For example, this can be achieved by decreasing the initial size $\sigma_{x0}$ of the atom cloud before releasing it into the dimple beam.
Alternatively, one could use more strongly elongated atom clouds before the first $\pi/2$ RF pulse is applied.

\par

By creating a non-expanding cold atom cloud in an elongated box-shaped potential, the cloud would become homogeneous, meaning that the errors in $\langle S_y(x)\rangle$ become comparable along the cloud and the magnetic field gradient can be easily calculated directly from Eq.~(\ref{grad1}), describing the non-expanding ($v_x=0$) case.
For example, this can be achieved by loading the atoms from a broad dipole trap directly into a single dimple beam, which is at both ends truncated with two narrow $532$~nm laser beams acting as repulsive barriers.
However, in this approach atomic diffusion will be present during the measurement protocol, which will substantially reduce the coherence time and the sensitivity of the instrument.
A possible solution is to prepare the atoms at much lower temperatures \cite{Higbie_2005}, which can be in principle achieved by evaporative cooling in such a box-shaped trap.

\par

In summary, we have demonstrated a simple and versatile method for measuring components of the magnetic field gradient in a single shot with an estimated absolute accuracy below $\pm0.2$~mG/cm ($\pm20$~nT/cm).
This method can be adopted to the majority of cold-atom setups with different atomic species, where it can serve as a
quantitative characterization tool or for the cancellation of magnetic field gradients \cite{Koschorreck_2011,Sycz_2018,Dedman_2007,Smith_2011}.
We emphasize that the sensitivity of our magnetic gradiometer suffers neither from atomic diffusion nor from fluctuations or drifting of homogeneous magnetic field since only spatially-dependent components of the magnetic field (the gradients and higher derivatives) contribute to the measured space modulated magnetization along the elongated cold atom cloud.
In addition, the presented method has the potential for miniaturization and for further improvements of its sensitivity to
magnetic field gradients in a single shot. 

\begin{acknowledgments}
We thank Wojciech Gawlik, Alan Gregorovi\v{c}, Matja\v{z} Gomil\v{s}ek and Philipp Haslinger for their comments and discussions.
This work was supported by the Slovenian Research Agency (research core fundings No. P1-0125 and No. P1-0099, and research project No. J2-8191).
\end{acknowledgments}


\end{document}